\documentstyle[psfig]{mn}
\newcommand{\ang}{\thinspace\hbox{\AA}}
\newcommand{\kms}{\thinspace\hbox{$\hbox{km}\thinspace\hbox{s}^{-1}$}}
\newcommand{\pc}{\thinspace\hbox{pc}}
\newcommand{\msun}{\thinspace\hbox{${\rm M}_{\odot}$}}

\begin{document}

\title[Kinematics of the helium accretor GP~Com]{Kinematics of
the helium accretor GP~Com}

\author[T. R. Marsh]
{T. R. Marsh\\
Department of Physics and Astronomy, University of Southampton, 
Highfield, Southampton SO17 1BJ}
\date{Accepted ??
      Received ??
      in original form ??}
\pubyear{1998}

\maketitle

\begin{abstract}
We present and analyse time-resolved $B$-band spectra of the
double-degenerate binary star, GP~Com. The spectra confirm the
presence and period ($46.5\min$) of the `S'-wave feature found by
Nather, Robinson and Stover. GP~Com is erratically variable at X-ray
and UV wavelengths. We have found the equivalent variability in our
data, which, as also seen in UV data, is mostly confined to the
emission lines. The HeII~4686 changes by the largest amount,
consistent with X-ray driven photo-ionisation. The flaring part of the
line profiles is broader than the average, as expected if they are
dominated by the inner disc. The HeII~4686 profile is especially
remarkable in that its blue-shifted peak is $1400\kms$ from line
centre compared to $700\kms$ for the HeI lines (the red-shifted peak
is blended with HeI~4713). We deduce that HeII~4686 emission is
confined to the inner 1/4 of the disc. We suggest that the activity of
the inner disc indicates that accretion is significant (and unstable)
there, in contrast to quiescent dwarf novae, in support of models in
which GP~Com is in a (quasi-)steady-state of low mass transfer rate.

GP~Com shows triple-peaked lines profiles which consist of the usual
double-peaked profiles from a disc plus a narrow component at line
centre. The latter has previously been ascribed to emission from a
nebula, although none could be found in direct images. However, we
find evidence for both radial velocity and flux variability in this
component, inconsistent with a nebula origin. The radial velocity
amplitude of $10.8\pm1.6\kms$ and its phase relative to the `S'-wave
are consistent with an origin on the accreting white dwarf, if the
mass ratio, $q = M_2/M_1$, is of order $0.02$, as expected on
evolutionary grounds. However this explanation is still not
satisfactory as the systemic velocity of the narrow component shows
significant variation from line to line, and we have no explanation
for this.

\end{abstract}


\section{Introduction}
GP~Com ($\equiv$G61--29) is a high proper motion star
\cite{Giclas1961} with an optical spectrum dominated by broad HeI
emission lines \cite{Burbidge1971}. Warner \shortcite{Warner1972} and
Smak \shortcite{Smak1975} found rapid photometric variability, similar
to the flickering observed in cataclysmic variable stars and suggested
that, like them, it is a binary star, although they found no
convincing periodicity.

Nather, Robinson \&\ Stover (1981, hereafter NRS) found a narrow
emission-line component which varied in radial velocity on a period of
$46.52\min$, providing strong support for the binary nature of
GP~Com. The variable component moved between the two outer peaks of
the triple-peaked emission lines, similar to phenomena observed in
hydrogen-dominated cataclysmic variable stars (e.g. U~Gem). Thus, by
analogy, NRS identified the moving component with the region where the
gas stream hits the accretion disc. The short period and absence of
hydrogen qualifies GP~Com as one of the AM~CVn group of accreting
double-degenerate systems; it is in fact the longest period of these
systems and stands out from the crowd as the only one with a spectrum
strongly in emission and as the only one to show an `S'-wave.

The most prominent element other than helium is nitrogen in the form
of NV in IUE \cite{Lambert1981} and HST \cite{Marsh1995} spectra and
NI in spectra covering the R- and I-bands \cite{Marsh1991}.  The
absence of hydrogen and the strong helium and nitrogen emission are
consistent with our seeing material from the core of the star that has
undergone hydrogen burning and CNO-cycle processing of most of the
carbon and oxygen into nitrogen. This is as expected given the very
short period of GP~Com.

The emission lines from GP~Com show the characteristic double-peaked,
broad profiles of emission from an accretion disc. However, unlike
hydrogen-dominated cataclysmic variables with such profiles, GP~Com
also sports a narrow component at the centre of each emission line
which we will refer to as the ``central spike''. The origin of this
component is unclear. NRS could find no radial velocity variability of
the central spike and suggested that it might come from a nebula
surrounding GP~Com, by analogy with old novae. However a subsequent
search failed to find any nebula \cite{Stover1983}. On the other hand
the mass donor star in GP~Com is probably of such a low mass that the
spike could originate on the mass accretor and still show little
radial velocity variability. We look at this component again in this
paper.

Another remarkable feature of GP~Com was seen in HST spectra
\cite{Marsh1995} in which flaring was seen in the emission lines,
representing a factor of 5 change in flux from minimum to
maximum. This is probably driven by X-ray variability
\cite{vanTeeseling1994} which might also influence the optical
lines. We look for this in the time-resolved data we present here.

We begin with a description of the observations and an analysis of
orbital variability in our data.

\section{The Observations and their Reduction}
We used the $2.5$m Isaac Newton Telescope on the island of La Palma to
take 414 spectra on three nights from the 21st to the 23rd April 1988.
The spectra were taken with the Intermediate Dispersion Spectrograph
(IDS) with an Image Photon Counting System (IPCS) detector, and cover
the range $4200$ to $5200$\AA\ in 2040 pixels with a full width half
maximum (FWHM) resolution of 2 pixels. Apart from the first six
exposures of 120 seconds each, all exposures were 100 seconds long,
taken through a $1.1$ by 50 arcsecond slit aligned to a position angle
of $4.7^\circ$ to capture simultaneous spectra of the variable GP~Com
and a nearby comparison star.  Arc spectra were taken every half hour
or so to track flexure in the spectrograph. Spectra of the comparison
star and the flux standards BD$+253941$ and Feige~34 
were taken to provide flux calibration.

The sky background was estimated by interpolation from uncontaminated sky 
regions on each side of the object spectra. After subtraction of the
background, the spectra were extracted by summation perpendicular to the 
dispersion. Flat field corrections were applied at the same time.

We fitted 13th order polynomials to the average arc spectrum for each 
night. The two lowest order terms were then refitted for the individual
arc spectra. The root mean square (RMS) of the fits was typically 
$0.07$\ang\ , equivalent to $5\kms$ or $1/7$th of a pixel at our dispersion.
The wavelength scales for GP~Com were then interpolated between
neighbouring arc spectra. This procedure eliminates the effects of flexure 
which amounted to $0.5$\ang\ during the night.

Finally the flux standard spectra were used to correct for the instrumental
sensitivity and the comparison star spectra were used to correct for
slit losses.

\begin{figure}
\hspace*{\fill}
\psfig{file=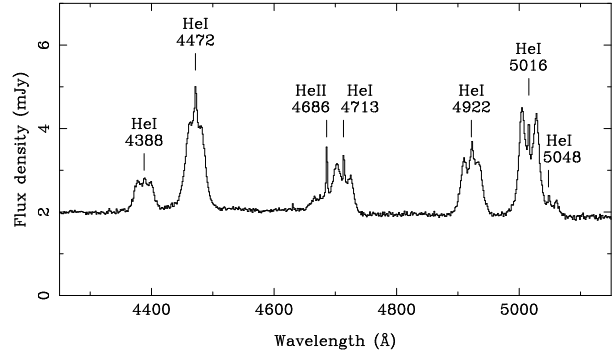,width=80mm}
\hspace*{\fill}
\caption{The average spectrum of our 414 spectra of GP~Com with the
central wavelengths and identifications of the most prominent lines
indicated.}
\label{fig:avspec}
\end{figure}

\section{Results}
\subsection{The Average Spectrum}
The average of our spectra has already been discussed in 
Marsh et al. \shortcite{Marsh1991}. For convenient 
reference, we show it once more in Fig.~\ref{fig:avspec}.
The spectrum confirms the earlier investigations of Burbidge \&
Strittmatter \shortcite{Burbidge1971} and NRS.  The only strong lines
detected are those of HeI and HeII, and these are broad and 
triple-peaked, a structure seen most clearly in the HeI 5015 line.

Smak \shortcite{Smak1975} and NRS suggest that the lines are made up
of a broad, double-peaked profile from an accretion disc plus a
separate narrow component near the centre of the line.  The variation
in the relative strength of the central ``spike'' from line to line
supports this hypothesis. Double-peaked profiles are well known from
the hydrogen-dominated cataclysmic variables \cite{Honeycutt1987}, 
but the central spike is unique to GP~Com.

\subsection{The Orbital Period}
The first reliable detection of a periodicity in GP~Com was made by
NRS who discovered a narrow emission-line component (in addition to
the central spike) which varied in velocity with a period of
$46.52\pm0.02\min$. The velocity of the component varied from $-670$
to $+670\kms$ taking it between the two outer peaks of the lines.
Such `S'-waves are well-known from other CVs, and are associated with
the region where the gas stream from the secondary star hits the
disc. If this is the case in GP~Com, then $46.52\min$ is also the
orbital period of the binary.

\begin{figure*}
\hspace*{\fill}
\psfig{file=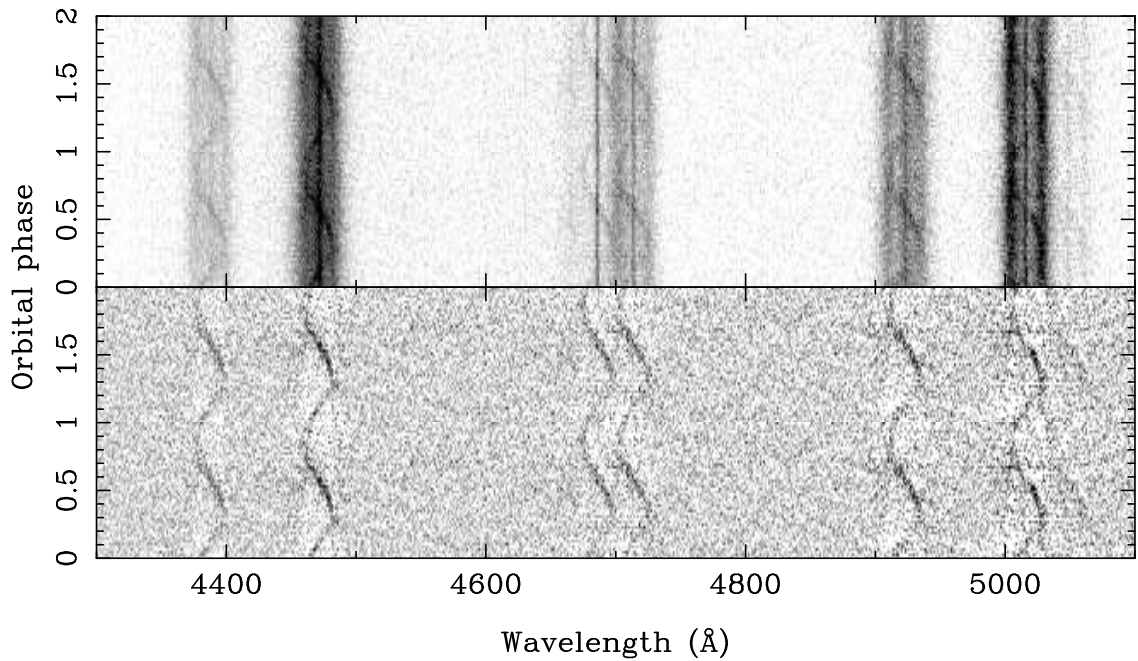,width=150mm}
\hspace*{\fill}
\caption{The upper panel shows the trailed spectrum of GP~Com
phase-folded and with the same cycle repeated twice. The continua of
the spectra are normalised to 1 and the spectra are displayed from a level
of 1 to $2.5$. In the lower panel the same spectra are shown after
subtraction of the average and are displayed from $-0.1$
to $+0.5$.}
\label{fig:trail}
\end{figure*}

The `S'-wave is also present in our data. In the upper panel of
Fig.~\ref{fig:trail} 
we show the phase binned trailed spectrum of GP~Com, and in the lower
panel we show the result of subtracting the average spectrum from
this. Fifty phase bins were equally spaced around the cycle and they
are repeated twice for clarity. The `S'-wave is clear in all the
lines, while there is little other variability. Fig.~\ref{fig:trail}
also shows the lack of radial velocity variation in the central spike.
Thus it is the `S'-wave that gives the best way of measuring the
orbital period in GP~Com.

In order to measure the period quantitatively we undertook a
multi-gaussian fit which was made to all the spectra and all lines
at once, representing the `S'-wave as a single gaussian with the same
sinusoidal radial velocity curve for all the lines. The period was
one of the free parameters of the fit, and is determined by the need
to keep the `S'-wave in step over the three nights of our run. We
obtained a period of $0.0323386 \pm 0.000002$ days or
$46.567\pm0.003\min$. This differs from NRS's value by $2.3\sigma$,
but given uncertainties caused by the round-off implicit in NRS's
value, we doubt this is significant.

There is no clear indicator of the conjunction phase for GP~Com, and
thus we chose an arbitrary zero point close to the middle of our
observations to compute phases. Thus we use the following ephemeris
for all phases in this paper:
\[ {\rm HJD} = 2447274.7 + 0.0323386 E . \]
This is the ephemeris used in constructing Fig.~\ref{fig:trail}.
We will discuss a possible constraint on the conjunction phase in 
section~\ref{spike}.

\subsubsection{Superhumps?}
The sub-class of cataclysmic variables called the SU~UMa stars are
known for quasi-periodic flaring behaviour during outburst which
occurs on a period somewhat longer than the orbital period. These
flares, usually known as ``superhumps'', are thought to be the result
of a distortion of the outer disc induced by the 3:1 resonance
\cite{Whitehurst1988} between the disc and binary orbit.  The
distortion only occurs if the 3:1 resonance radius is inside the disc
and this requires the mass ratio $q$ ($= M_2/M_1$) to be less than
about $0.25$. This neatly explains why the SU~UMa stars are all at the
short period end of the cataclysmic variable period distribution
because then the secondary star is of relatively low mass. The
restriction on mass ratio is easily satisfied by the even shorter
period GP~Com, and so it is of interest to see whether it shows any
superhump-like behaviour. Since our spectra are slit-corrected we can
attempt this.

We computed light curves over two regions avoiding the emission
lines. The first of these covered 4250 to 4350 and 4510 to 4640\ang\,
while the second covered 4750 to 4880 and 5080 to 5190\ang.
\begin{figure}
\hspace*{\fill}
\psfig{file=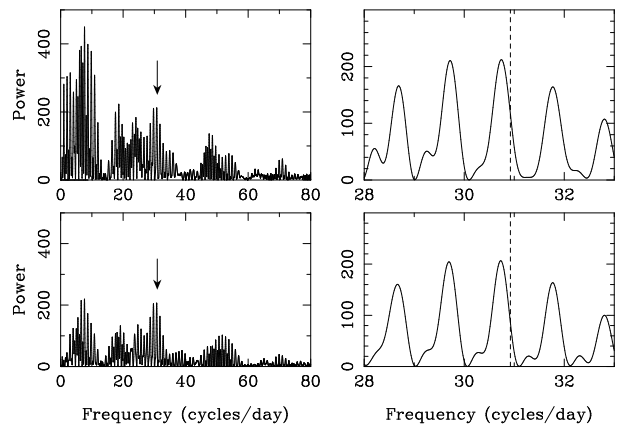,width=80mm}
\hspace*{\fill}
\caption{The periodograms of the light curves covering the red and
blue halves of our data (excluding the emission lines) are plotted in 
the lower and upper panels respectively. The right-hand panels
shows a detailed view near the spectroscopic frequency which is
indicated by the dashed lines on the right and the arrows on the left.}
\label{fig:perg}
\end{figure}
\begin{figure*}
\hspace*{\fill}
\psfig{file=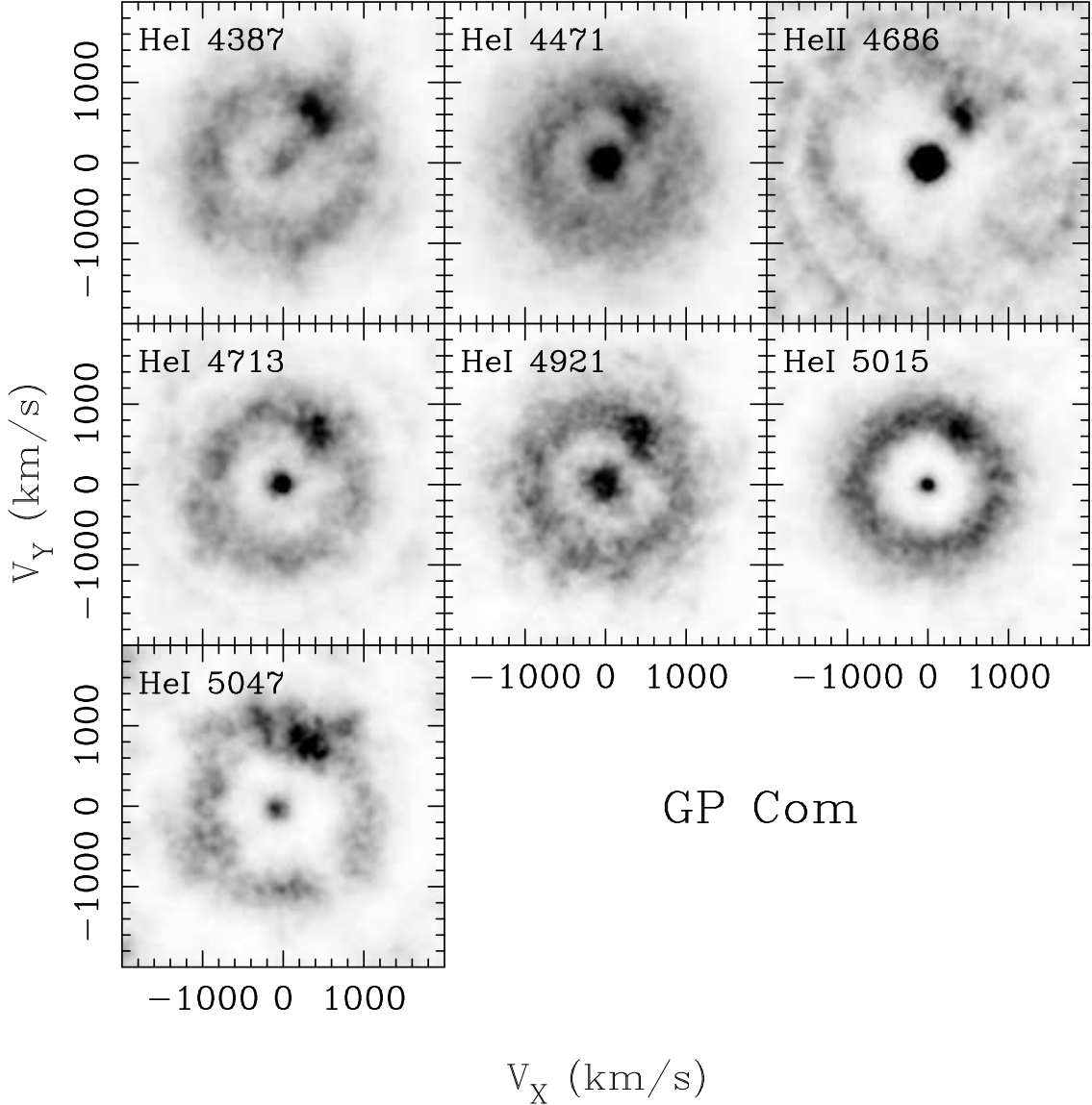,width=150mm}
\hspace*{\fill}
\caption{The Doppler images of 7 lines of GP~Com. The arbitrary
zero-point of our ephemeris means that these images are subject to
an unknown rotation compared to the standard coordinate system
used in displaying such images.}
\label{fig:dopp}
\end{figure*}
Periodograms \cite{Scargle1982} of these light curves are shown in
Fig.~\ref{fig:perg}. These do indeed show some evidence for a signal
on a slightly longer period than the spectroscopic one; the difference
visible in the right-hand panels of Fig.~\ref{fig:perg} corresponds to
a difference of about $0.4$ cycles over the 2 day baseline of our
observations, which is easily measurable. Thus our data
are at least consistent with the possibility of superhump-like
behaviour in GP~Com. However, given that the signal in the continuum 
near the spectroscopic period is only one of several peaks of similar
power, the evidence can only be regarded as suggestive. It should be
noted that superhumps in SU~UMa stars are only seen during some
outbursts, and so it would not be surprising if GP~Com failed to show them.

\subsection{Doppler Images}
\label{sec:dopp}
The emission lines in GP~Com show the broad, Doppler-shifted profiles 
produced by accretion discs. Doppler tomography provides a powerful
method of interpreting such profiles. We imagine that the profiles are
built from the sum of many `S'-waves which have radial velocity
\[ V(\phi) = \gamma-V_X \cos 2\pi \phi + V_Y \sin 2\pi \phi ,\]
at orbital phase $\phi$. The strength of each `S'-wave is plotted with
coordinates $V_X$, $V_Y$ to build up an image. The coordinates are chosen for
consistency with a coordinate system in which the $X$ axis points
from the primary to the secondary star and the $Y$ axis points in the
direction of the secondary star's motion. A detailed description of
the computation of the images is given in Marsh \& Horne
\shortcite{Marsh1988}.

Images of the seven lines are presented in Fig.~\ref{fig:dopp}. These
images were computed in three sets, the first covering HeI 4387 and
4471, the second HeII 4686 and HeI 4713 and the third HeI 4921, 5015
and 5047. In this way blending between lines could be accounted for.
The fits to the lines were very good with reduced $\chi^2$ of order
$0.8$.

All lines show a similar structure with a central spot, matching the
central spikes of Figs.~\ref{fig:avspec} and \ref{fig:trail}, an
offset spot equivalent to the `S'-wave and a smooth ring-shaped
background for the rest of the disc. We have scaled the lines so that
the ring from the disc is seen at about the same level in each case
which brings out the marked variation in the relative strengths of the
central spike and bright-spot components.

The outer ring visible in HeII~4686 is a consequence of its
blending with HeI~4713 (which has a similar but less obvious
artefact). It is the inner ring that corresponds to the edge of
the HeII 4686 emission region and it can be seen to be markedly
larger than its equivalents in the HeI lines. This indicates that the
HeII 4686 emission does not cover all of the disc, a point we will
return to in section~\ref{sec:flare} when we discuss flaring behaviour
in GP~Com.

\subsection{The `S'-wave}
\label{sec:swave}
\begin{figure}
\hspace*{\fill}
\psfig{file=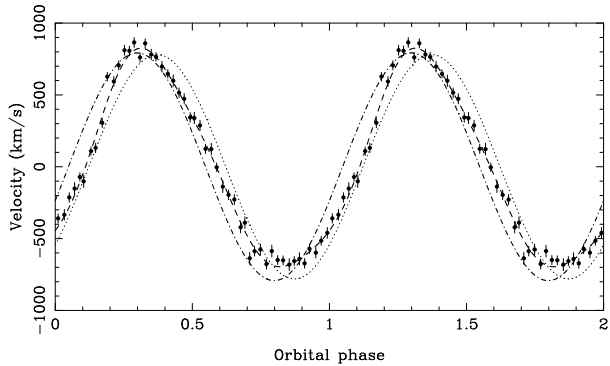,width=80mm}
\hspace*{\fill}
\caption{The figure shows the `S'-wave velocity measured from all
lines simultaneously for 50 phase-folded spectra (repeated over two
cycles). The dashed line shows a fit with the first three Fourier
coefficients included. The dash-dotted and dotted lines are the
velocities of the stream and disc at the bright-spot for mass ratio
$q = 0.02$ and a disc extending out to 60\% of the Roche lobe (see
section~\protect\ref{sec:spike}).}
\label{fig:swave}
\end{figure}

The spot equivalent to the `S'-wave appears to have some structure in
several of the images of Fig.~\ref{fig:dopp}, most obviously in HeI
4387.  On examining the data more closely (Fig.~\ref{fig:trail}), it
turns out that the `S'-wave is significantly non-sinusoidal, a
phenomenon also seen in the strong `S'-wave of WZ~Sge (this is best
seen by looking at Fig.~\ref{fig:trail} sideways). In order to
assess the departure from a sinusoid, we returned to the multi-gaussian
fits, forcing the velocity of the S-wave to be the same in all
lines. We first fitted the `S'-wave velocity as before when
determining the period but now, rather than just a simple sinusoid, we
fitted a Fourier series of the form:
\[ V(\phi) = - \sum_{n=1}^3 A_n \cos 2\pi n \phi + \sum_{n=1}^3 B_n
\sin 2\pi n \phi .\]
With this as a starting point, we then measured velocities
individually from 50 phase-folded spectra (the signal-to-noise being
too low to attempt this on the raw data).
The results and Fourier fit are plotted in Fig.~\ref{fig:swave}.
The parameters of the Fourier fit are listed in
Table~\ref{tab:fourier}. 
\begin{table}
\caption{Parameters of the Fourier fit to the S-wave in GP~Com}
\begin{tabular}{lr@{$\,\pm\,$}lr@{$\,\pm\,$}lr@{$\,\pm\,$}l}
Coefficient      &  \multicolumn{2}{c}{1} &  \multicolumn{2}{c}{2} & 
\multicolumn{2}{c}{3} \\[1mm]
                 & 
\multicolumn{2}{c}{${\rm km}\,{\rm s}^{-1}$} & 
\multicolumn{2}{c}{${\rm km}\,{\rm s}^{-1}$} & 
\multicolumn{2}{c}{${\rm km}\,{\rm s}^{-1}$} \\
Cosine amplitude &$402$&$7$ & $49$ &$7$ & $-7$&$7$ \\
Sine amplitude   &$631$&$7$ & $-42$&$7$ &$-44$&$7$ 
\end{tabular}
\label{tab:fourier}
\end{table}
Both figure and table exhibit the substantial departure from a pure
sinusoid.  Non-sinusoidal velocity changes of this manner violate one
of the basic assumptions of Doppler tomography, but are easily
produced by phase-dependent visibility in the emission regions.  For
example, at one phase we may see emission from the stream, whereas at
another emission from the disc near the stream may be easier to
see. In general these will have different velocities and thus we see a
non-sinusoidal behaviour. Such visibility variations imply that
vertical structure in the disc or stream plays a significant role in
GP~Com, suggesting that it is of high orbital inclination.  The
Doppler maps cannot reflect this complexity and instead we see an
averaged view of the system with emission spread over all contributing
sites.  This explanation is nicely supported by the phase-dependent
flux variation in the `S'-waves of Fig.~\ref{fig:trail}.  Similar
behaviour can be seen in the `S'-wave of the hydrogen-dominated
system, WZ~Sge \cite{Spruit1998}.

To illustrate the difference between the stream and disc velocities we
have plotted the `S'-waves predicted for stream only (dash-dotted
line) and disc only (dotted) in Fig.~\ref{fig:swave}. The predicted
velocities are based upon a mass ratio of $q = 0.02$ and a disc of
radius $0.6\,{\rm R}_{\rm L1}$. Ideally, the observed velocities
should be bracketed by these two paths. This is near enough the case
for the above explanation to be credible. The phasing and scaling used
in producing the prediction is discussed in the next section.

\begin{table}
\caption{S-wave parameters of each line.}
\begin{tabular}{lr@{$\,\pm\,$}lr@{$\,\pm\,$}l}
Line      &   \multicolumn{2}{c}{$V_X$} & 
\multicolumn{2}{c}{$V_Y$}         \\[1mm]
          & \multicolumn{2}{c}{${\rm km}\,{\rm s}^{-1}$}& 
\multicolumn{2}{c}{${\rm km}\,{\rm s}^{-1}$} \\
HeI 4387  &    $412$ & $18$ & $614$ & $18$ \\
HeI 4471  &    $384$ & $15$ & $630$ & $15$ \\
HeII 4686 &    $432$ & $19$ & $582$ & $18$ \\
HeI 4713  &    $425$ & $20$ & $658$ & $20$ \\
HeI 4921  &    $400$ & $16$ & $598$ & $16$ \\
HeI 5015  &    $385$ & $16$ & $658$ & $16$ \\
HeI 5047  &    $330$ & $51$ & $617$ & $51$ 
\end{tabular}
\label{tab:swave}
\end{table}
Measurements of the amplitudes for the individual lines are listed in
Table~\ref{tab:swave}. In this case a simple sinusoid has been fitted
to each line. We will consider these further in the next section.

\subsection{The Central Spike}
\label{sec:spike}
The absence of obvious radial velocity variation in the central spike
led to suggestions that it arose in a nebula, which however has not
been found \cite{Stover1983}.  However, although the absence of
variations rules out the mass donor, the disc and the bright-spot, as
we pointed out earlier, the accreting white dwarf remains a
possibility because the extreme mass ratio expected of GP~Com (see
section~\ref{sec:param}) means that its semi-amplitude is only of order
$10\kms$. Therefore a measurement of the radial velocity amplitude of
the central spike provides an important constraint upon its origin.

Actually measuring the amplitude is difficult.  In principle it can be
measured by determining the position of the central spot in the
Doppler images. However, this is adversely affected by the systemic
velocity shifts visible in the mean spectra (which tend to blurr the
central spot) and it is also difficult to estimate uncertainties with
this method. On the other hand one can't simply measure velocities
directly without accounting for the presence of the
`S'-wave. Therefore we again applied the multi-gaussian fitting
method. With the spike and `S'-wave modelled as gaussians, and with
velocities fitted by the usual $\gamma - V_X \cos 2\pi \phi + V_Y\sin
2\pi \phi$ function, the fitted parameters are listed in
Table~\ref{tab:spike}.
\begin{table}
\caption{Velocity parameters of the central spikes.}
\begin{tabular}{lr@{$\,\pm\,$}lr@{$\,\pm\,$}lr@{$\,\pm\,$}l}
Line      &  \multicolumn{2}{c}{$\gamma$} &  \multicolumn{2}{c}{$V_X$}
& \multicolumn{2}{c}{$V_Y$}         \\[1mm]
          & \multicolumn{2}{c}{${\rm km}\,{\rm s}^{-1}$}& 
\multicolumn{2}{c}{${\rm km}\,{\rm s}^{-1}$}& \multicolumn{2}{c}{${\rm
km}\,{\rm s}^{-1}$} \\
HeI 4387  &  $43.9$ & $9.3$         &   $8.6$ & $13.1$& $28.3$ & $13.1$ \\
HeI 4471  &  $29.7$ & $3.5$         &  $-0.5$ & $4.9$ & $-0.1$ & $4.9$ \\
HeII 4686 &  $22.8$ & $1.7$         &  $-9.5$ & $2.4$ & $-3.6$ & $2.4$ \\
HeI 4713  &  $42.5$ & $2.6$         & $-12.9$ & $3.7$ & $-4.7$ & $3.7$ \\
HeI 4921  &  $55.7$ & $4.6$         & $ -9.3$ & $6.5$ & $-1.9$ & $6.4$ \\
HeI 5015  &  $-3.4$ & $2.6$         & $-13.4$ & $3.7$ & $-5.8$ & $3.7$ \\
HeI 5047  &  $45.4$ & $9.3$         & $-43.1$ & $13.2$& $-25.7$& $13.2$ \\
All       &  \multicolumn{2}{c}{--} & $-10.3$ & $1.6$ & $-3.9$ & $1.6$ 
\end{tabular}
\label{tab:spike}
\end{table}
The last entry denoted by ``All'' was derived from a simultaneous fit
to all lines in which they were forced to have the same $V_X$ and
$V_Y$.  Individually the signal-to-noise in the lines is too marginal
to be confident of a detection, but the simultaneous fit to all of
them, which gives $V_X = -10.3 \pm 1.6\kms$, $V_Y = -3.9 \pm 1.6\kms$,
is statistically significant. However, it should be remembered that
this measurement was made against a background of disc and `S'-wave
and that the semi-amplitude is only of order 1/10$^{\rm th}$ of the
FWHM of the spike. Nevertheless the detection supports the accretor
origin for the central spike. 
\begin{figure*}
\hspace*{\fill}
\psfig{file=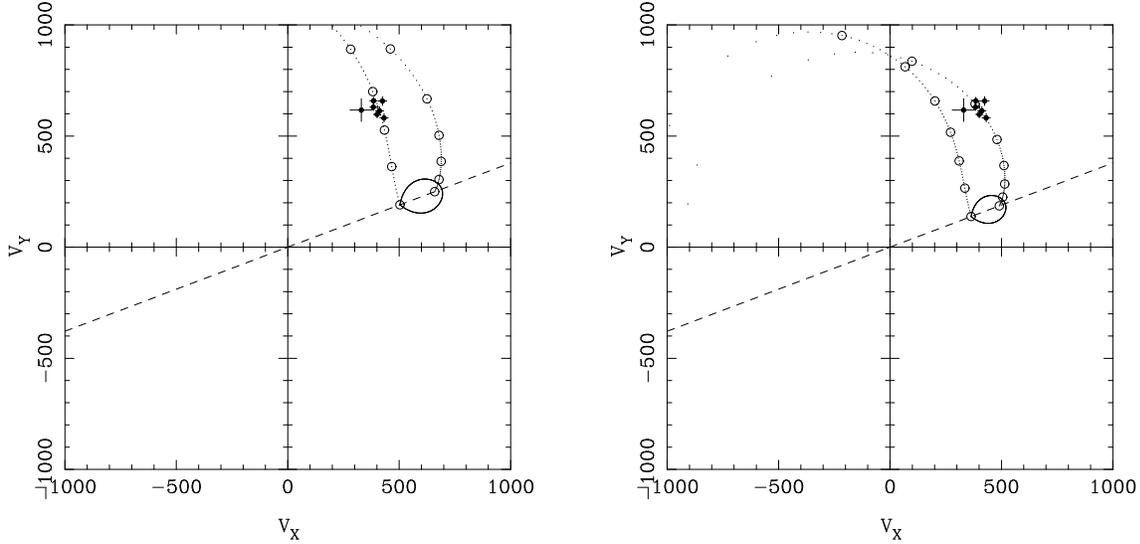,width=150mm}
\hspace*{\fill}
\caption{The two panels shows the measured `S'-wave points from
Table~\protect\ref{tab:swave}. Plotted over these are the predicted
velocities of the stream (left-hand dotted lines), the keplerian
velocity of the disc along the path of the stream (right-hand dotted
lines) and the secondary star (egg-shaped object), all scaled and
rotated to match the measured amplitude and phase of the central
spike, assuming that it tracks the primary star. In the left panel a
mass ratio $q = 0.017$ has been used while $q = 0.023$ has been used
in the right panel. Dots and circles indicate steps of $0.01$ and
$0.1\,{\rm R}_{\rm L1}$ in distance from the primary star.}
\label{fig:roche}
\end{figure*}

After making a small correction for noise-induced bias,
our measurement is equivalent to an amplitude of $10.8\pm1.6\kms$.
This can be compared to NRS's measurement of $14.3\pm3.7\kms$ based
upon the emission line wings. The values are consistent, although
unfortunately NRS do not specify the phasing of their measurement.
We have not tried to measure the amplitude from the line wings as
they are so broad that the measurement of such small amplitudes
cannot be relied upon, as NRS warn.

If the spike truly does reflect the motion of the accretor, and the 
`S'-wave originates from the gas stream/disc impact region, then 
their relative phases and amplitudes depend upon the mass ratio.
The connection is illustrated in Fig.~\ref{fig:roche} which shows that
the S-wave parameters listed in Table~\ref{tab:swave} (plotted as the
cluster of points in upper-right quadrant of each panel) are consistent
with the measured position of the spike (which lies on the straight,
dashed lines) for a mass ratio $q = 
 0.017$ if the `S'-wave represents the velocity of the gas stream or
$q = 0.023$ if it represents the velocity of the disc along the path
of the stream. In the former case the disc would extend to 75\% of the
way to the inner Lagrangian point, whereas in the latter it would only
need reach $\approx 50$\% of the way. Both mass ratios and disc radii
are reasonable for GP~Com, and provide further circumstantial evidence
that the spike may indeed be from the accretor. If this interpretation
is correct, then inferior conjunction of the secondary star (the
expected phase of eclipse of the disc, if there was one), occurs at
orbital phase $0.19\pm0.02$ on our ephemeris. The same method was used
to produce the predicted stream and disc `S'-waves plotted in 
Fig.~\ref{fig:swave}.

\subsection{Stochastic variability}
\label{sec:flare}
So far we have mainly considered variations with orbital
phase. However, strong flaring behaviour has been observed in HST data
which does not appear to be related to orbital phase \cite{Marsh1995}.
In the HST data the flux of NV 1240 was seen to increase by a factor
of up to five during three flares which occurred in the 13 hour
observing interval, while the continuum only increased by $\approx
40$\%. It is likely that this variation is driven by X-ray variability
through photo-ionisation, and indeed, variations have been seen in
ROSAT data on GP~Com \cite{vanTeeseling1994}. If so,
we can expect similar effects at optical wavelengths.
\begin{figure}
\hspace*{\fill}
\psfig{file=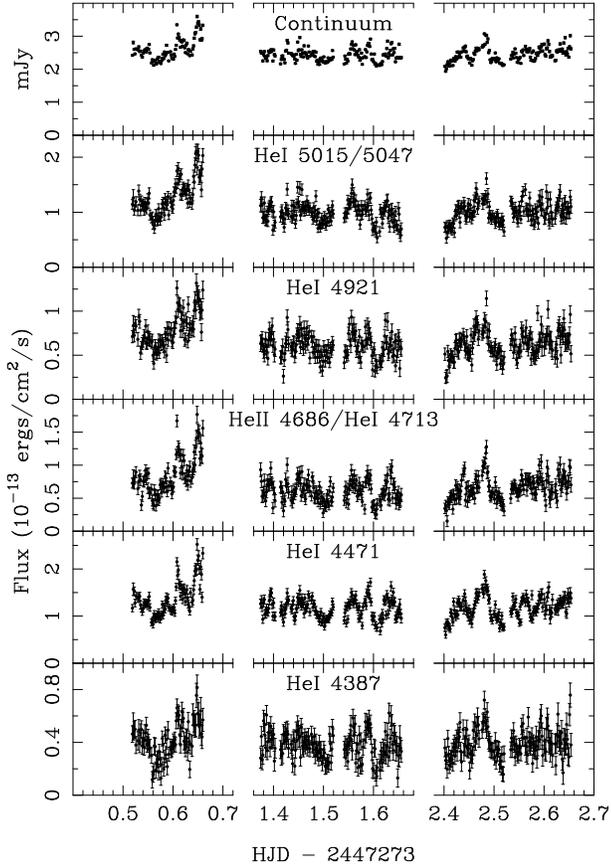,width=80mm}
\hspace*{\fill}
\caption{The figure shows the light curves of the emission lines and
continuum of GP~Com over the three nights of our run. Across the
figure each panel represents the same time interval.}
\label{fig:lig}
\end{figure}
That this is so is demonstrated in Fig.~\ref{fig:lig} which shows 
the light curves of the continuum and emission lines of GP~Com during 
our run. There is significant, correlated variability in all
components, and it has a larger amplitude in the lines than the continuum.

We can derive a spectrum for the flaring component using the method
employed by Marsh et al. \shortcite{Marsh1995}. In this the spectra
are modelled as the sum of a spectrum representing the mean plus
multiples of another spectrum representing the flaring component.
Both spectra and the multipliers (which are different for each of the
414 spectra) are optimised by minimising $\chi^2$.
\begin{figure}
\hspace*{\fill}
\psfig{file=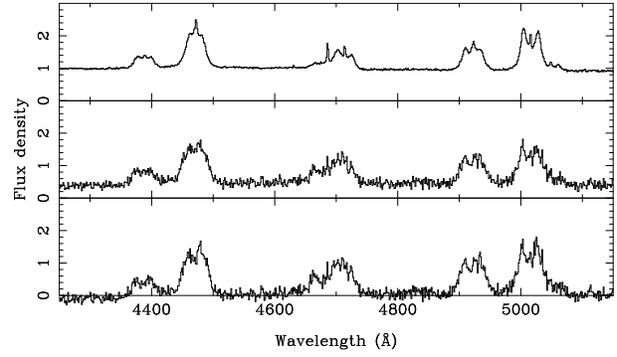,width=80mm}
\hspace*{\fill}
\caption{The panels show the mean GP~Com spectrum (top), the
flare spectrum derived from the raw spectra (middle) and the
flare spectrum derived after subtracting the continuum (bottom).
Each spectrum has been rebinned into 800 pixels.}
\label{fig:flare}
\end{figure}
The constant and flare components derived from our data are shown in
the top two panels of Fig.~\ref{fig:flare}. It has to be remembered
that our spectra were taken through a narrow slit, and although they
were corrected for slit losses, it is likely that some part of the
``flaring'' could be artificially induced. This would have the effect
of making the flare spectrum look like the mean spectrum.  Therefore
as a further check we repeated the computation of the flare spectrum
after subtracting the continuua of the spectra. This allows the
variations in the lines to determine the flare spectrum rather than
the lines and continuum and should weaken the influence of poor slit
corrections, although not eliminate them entirely. The result of this
is shown in the lower panel of Fig.~\ref{fig:flare} and probably gives
the most accurate representation of the line profiles.
\begin{figure}
\hspace*{\fill}
\psfig{file=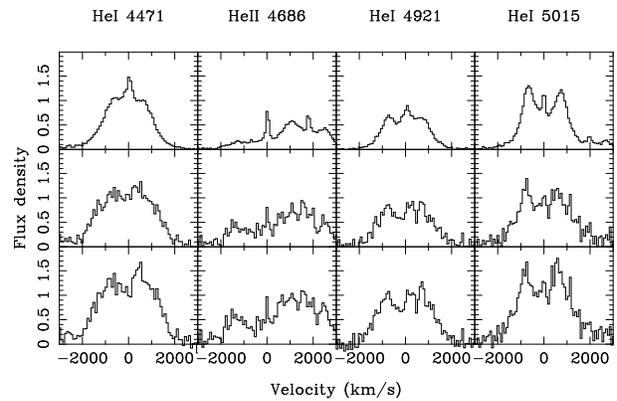,width=80mm}
\hspace*{\fill}
\caption{The panels show the same spectra as
Fig.~\protect\ref{fig:flare} but now concentrate upon the line
profiles. The continua have been subtracted from all three spectra.
Note that the HeII 4686 is blended with HeI 4713.}
\label{fig:flare_prof}
\end{figure}
Expanded views of four of the line profiles are shown in 
Fig.~\ref{fig:flare_prof}.

Looking first at the upper two panels confirms the behaviour seen in
the light curves of Fig.~\ref{fig:lig}. The emission lines are
relatively stronger than the continuum in the flare compared to the
mean.  This is exactly as found in the HST data \cite{Marsh1995},
although not to such a marked extent (the possible contamination due
to poor slit loss correction should be remembered however). The line
ratios in the two spectra are very similar with the exception of HeII
4686 which is stronger in the flare component; with hindsight the
HeII 4686/HeI 4713 light curve can be seen to be the most modulated in
Fig~\ref{fig:lig}. This supports photo-ionisation as the driver of the
flares. Finally there are substantial differences in the line profiles
(Fig.~\ref{fig:flare_prof}). The profiles in the flare spectrum have
broader wings than they do in the mean spectrum, indicative of a
larger contribution from the inner disc and they also have a weaker
central spike component compared to the mean. However, it is important
to note that the flare spectrum does appear to have some central spike
component, although the caveat about slit loss contamination is worth
repeating.

The weakening of the central spike leads to profiles with more
pronounced double-peaks.  Most remarkable of all is the HeII 4686
profile which in the flare spectrum has a blue-shifted peak at around
$-1400\,{\rm km}\,{\rm s}^{-1}$ (the blending with HeI 4713 rendering
it impossible to be sure of the red-shifted peak). If interpreted as
a standard profile from an accretion disc, this suggests that the
HeII~4686 emission region only extends out to about 1/4 of the radius
of the disc (assuming that the HeI lines come from all of it). This
peak is consistent with the radius of the inner ring of the HeII~4686
map of Fig.~\ref{fig:dopp}.

\section{Discussion}

\subsection{Origin of the Central Spike}
\label{spike}
The persistent narrow emission at the centre of all the lines is a
puzzling feature of GP~Com. The nearest equivalent we know of is the
zero velocity emission observed in IP~Peg and SS~Cyg
\cite{Steeghs1996}, but this was only seen during outburst. The
parameters in Table~\ref{tab:spike} provide several constraints upon
its origin. The amplitude of any radial velocity variations is low.
The spike is seen best in HeI 4713, HeI 5015 and HeII 4686 and these
lines indicate a narrow FWHM of $\approx 120 \kms$. We take these
lines as representative since the ambiguity about whether flux is from
the spike or from the disc is likely to cause much larger systematic
errors for the other lines.

Taken together, the small width and the radial velocity amplitude
allow only a few possible sites of origin which are (a) the accreting
white dwarf, (b) a nebula or (c) a wind or jet. Any other part of the
binary moves too fast (e.g.\ the donor star) or would produce too
broad a profile (e.g.\ the disc). All the above sites have their own
problems however.

A wind origin faces difficulties since the winds observed in normal
cataclysmic variables are too highly ionised to produce HeI emission
and have large velocities characteristic of the escape velocity of the
white dwarf \cite{Drew1987}.  It is hard to see how expansion
velocities of order $60\kms$ arise in such circumstances. Moreover a
wind origin has no good explanation for the variations in the systemic
velocities, most obvious in HeI~5015. 

The main point in favour of the accreting star is our detection of
radial velocity variation, albeit of low amplitude. As we showed in
section~\ref{sec:spike} the measured amplitude and phase of the spike
are in accord with the `S'-wave amplitude and phase. However, once
more the systemic velocities are hard to understand, although 
the primary star has the possibilities of gravitational, pressure
\cite{Koester1987} and Zeeman shifts. It is encouraging that the most
discrepant line, HeI~5015, is the only singlet line in our wavelength
range, but there does not seem to be enough known about these
lines for us to say anything more.

A nebula origin copes best with the systemic velocities.  HeI~5015 is
the line most sensitive to optical depth effects. Under case~A
conditions the HeI 5015/4922 ratio should be of order $0.1$ whereas
under case~B it increases to about $2.5$ \cite{Brocklehurst1972}. Thus
a correlation between the velocity of the emitting material and its
density could be the reason behind the HeI~5015 anomaly. The nebula,
if it exists, would have to be asymmetric, which is no great obstacle.
There are however other problems with a nebula origin. First, it is
not resolved. A nebula expanding at $60\kms$ would reach a size of 1
arcsecond in only 10 years if GP~Com lies at the upper limit of
$100\pc$ deduced by Marsh et al.\ \shortcite{Marsh1991}. We find no
evidence for any extended emission in our long slit spectra and
estimate that an extension of order 1 arcsecond should have been
detected; Stover \shortcite{Stover1983} found that what appeared to be
extended emission around GP~Com is in fact a group of stars. Perhaps
the most significant problems are the radial velocity and flux
variability that we have detected. The nebula explanation can only
survive these if they are dismissed as systematic artefacts. Given the
small amplitude of the radial velocity variations, it is hard to rule
out this possibility. However, while we cannot be sure that poor slit
loss correction has not artificially produced the spike component in
the flare spectra, our feeling is that it is too strong for this to be
the case. Luckily, it should be possible to improve upon this with
better data in the future.

\subsection{Parameters of GP~Com}
\label{sec:param}
The evolution that leads to a system such as GP~Com has been discussed
in several papers \cite{Savonije1986}. At any particular orbital period there
are two possibile configurations. In one case the donor may be a
helium star of relatively high mass and luminosity. In this case the
orbital period decreases with time. Alternatively, at a later stage after the
system has passed its minimum period (of order $10\min$), the donor
adopts a semi-degenerate structure of very low mass and luminosity.
In this state as the donor loses mass it increases in radius and the
orbital period increases as well.  It is unable to reach a truly
degenerate structure because its thermal timescale becomes so
long. The latter case is almost certainly the one that applies to
GP~Com since there is no sign of the donor star. Our favoured mass
ratio of $q = M_2/M_1 \approx 0.02$ also supports this scenario. Based
upon Savonije et al.'s work, Warner \shortcite{Warner1995} deduces
that
\[ M_2 = 0.0186 P^{-1.274} \msun ,\]
where the orbital period $P$ is measured in hours. This gives a mass
of order $0.026\msun$ for the donor star. Given an accretor mass of
order $1\msun$, this is acceptably close to the mass ratio we deduce,
and confirms the status of GP~Com. Using the relation given by Warner
for a fully-degenerate secondary (his equation 9.43) leads 
to $M_2 = 0.009\msun$, still consistent with our mass ratio if the 
accretor is more like $0.5\msun$.

For $q = M_2/M_1 = 0.02$, $K_1 = 10.8\kms$ and $P = 46.567\min$, we
calculate $M_1 \sin^3 i = 0.55\msun$, a reasonable
value. Uncertainties in $q$ and $K_1$ are too large to deduce any
useful value of the inclination however, and we just note again that
the non-sinusoidal `S'-wave behaviour suggests that it might be high,
although there is no sign of an eclipse.

\subsection{Flaring in GP~Com}
The flaring we have found at optical wavelengths, and seen more
dramatically still at UV wavelengths, is another unusual feature of
GP~Com. It strongly suggests that the inner disc is variable in a way
not seen in the systems that look most like GP~Com in terms of their
optical spectra, the quiescent dwarf novae (leaving aside the absence
of hydrogen!). GP~Com has never been seen to show outbursts, and there
has been discussion as to whether this is because it has a very long
inter-outburst interval or because it is in a steady-state of very low
mass transfer rate \cite{Warner1995}. The former possibility is quite
reasonable given the decades-long intervals between the outbursts of
the short-period dwarf novae known as the WZ~Sge stars. On the other
hand, it has long been realised that, if the thermal instability model
of dwarf nova outbursts is correct, a disc composed largely of helium
may behave very differently from one dominated by hydrogen
\cite{Smak1983,Tsugawa1997}. The turning points on the thermal
equilibrium curve plotted as $\log T$ versus $\log \Sigma$ (the
``S-curve'') are at a higher temperature and surface density for
helium compared to hydrogen \cite{Tsugawa1997}.  As a result, it is
more likely that the entire disc can be on the lower branch of the
S-curve, and thus the system will not undergo outbursts.

We suggest that the high-level of X-ray to optical flux in GP~Com
\cite{Verbunt1997} and the flaring behaviour support the latter
possibility. That is, matter accretes at all radii in the disc rather
than accumulating in the outer disc as it is believed to do in
quiescent dwarf novae.  In this state the disc can be optically thin
\cite{Tsugawa1997}, which explains the strong emission lines displayed
by GP~Com, in contrast to the rest of the AM~CVn group.  Of course,
some other instability must be occurring in the inner disc to explain
the flaring; its cause remains to be determined.

\section{Conclusions}
We have analysed time-resolved spectrophotometry of the
double-degenerate binary GP~Com. We confirm the presence of the
`S'-wave feature found by NRS and refine its period to
$46.567\pm0.003\min$. We have detected a small radial velocity
variation with a semi-amplitude of $10.8\pm1.6\kms$ in the sharp
component at the centre of the emission lines, which may indicate that
it comes from the accreting primary star. The amplitude and phase are
consistent with such an origin together with the measured parameters
of the `S'-wave if the mass ratio $q = M_2/M_1$ is of order $0.02$.
This is roughly as expected if GP~Com is now increasing in period
and has a near-degenerate donor star. However, the systemic velocity
of the narrow component, which varies from line to line, remains to be
explained.

While the orbital variations that we find are consistent with earlier
data, we have also discovered erratic flaring which we believe to be
analogous to similar but more obvious behaviour seen at UV
wavelengths.  We suggest that this supports models in which GP~Com's
disc is on the lower branch of the thermal instability curve and does
not undergo the global outbursts followed by intervals of mass
accumulation as occur in dwarf novae.

\section*{Acknowledgements}
TRM was supported by a PPARC Advanced Fellowship during the course of
part of this work.  The data reduction and analysis were carried out on the
Southampton node of the UK STARLINK computer network.

\end{document}